\documentclass[particles,article,submit,moreauthors,pdftex,10pt,a4paper]{Definitions/mdpi} 

\firstpage{1} 
\pubvolume{xx}
\pubyear{2019}
\copyrightyear{2019}
\history{Received: date; Accepted: date; Published: date}

\Title{Using the Beth-Uhlenbeck approach to describe the kaon to pion ratio in a 2+1 flavor PNJL model}

\Author{D. Blaschke $^{1,2,3}$, A. Friesen $^{2}$, Yu.~Kalinovsky$^{2}$ and A. Radzhabov $^{4,5}$}

\AuthorNames{D. Blaschke, A. Friesen, Yu.~Kalinovsky and A. Radzhabov}

\address{%
$^{1}$ \quad Institute for Theoretical Physics, University of Wroc{\l}aw, 50-204 Wroc{\l}aw, Poland\\
$^{2}$ \quad Joint Institute for Nuclear Research, Dubna, 141980 Dubna, Russia\\
$^{3}$ \quad National Research Nuclear University (MEPhI), 115409 Moscow, Russia \\
$^{4}$ \quad Matrosov Institute for System Dynamics and Control Theory, Irkutsk 664033, Russia \\
$^{5}$ \quad  Irkutsk State University, Irkutsk 664003, Russia \\
}

\corres{Correspondence: aradzh@icc.ru}

\abstract{The  kaon to pion ratios are discussed in the framework of a 2+1 flavor PNJL model.
In order to interpret the behaviour of bound states in medium the Beth-Uhlenbeck approach  is used.
It is shown that in terms of phase shifts in the $K^+$ channel an additional low-energy mode could appear as a bound state  in medium since the  masses of the quark constituents are different. The comparison with experimental data for the ratios is performed and the influence of the anomalous mode to the "horn" effect in the $K^+/\pi^+$ ratio is discussed.
}


\preto{\abstractkeywords}{\nolinenumbers}

\begin{document}
\section{Introduction}
The asymptotic freedom feature of quantum chromodynamics (QCD) results in the behavior of quarks and gluons as point-like particles with rather weak interactions at high energy transfer. 
Similarly, at very high temperatures or/and chemical potentials the quark-gluon plasma (QGP) could be formed. 
However, the most interesting region of the QCD phase diagram is subject to nonperturbative effects like bound state and condensate formation since the strong coupling constant is not a small parameter. 
Probably, the only method to calculate nonpertubative observables  from "first principles" is to perform numerical simulations in the lattice QCD formulation. 
Despite recent progress the applicability of lattice QCD calculations is still limited to the low-density region of the QCD phase diagram. 
Moreover, the interpretation of numerical results in order to theoretically understand the underlying mechanisms is desired.
Therefore, in absence of "first principle" results one can use effective models based on QCD symmetries. 
The Nambu--Jona-Lasinio model based chiral symmetry and its spontaneous breaking is widely used for investigations of the phase diagram. 
The extension of the NJL model by an additional coupling of the quarks to background gauge degrees of freedom on the basis 
of the Polyakov loop potential (or PNJL model \cite{Ratti:2005jh,Roessner:2006xn}) introduces the finite temperature aspect of confinement, i.e. removes the contribution of free quarks to thermodynamic observables. 
In search for the possible formation of new phases of strongly interacting matter one needs a process or observable which can serve as an indicator. 
The puzzling observation of an enhancement of the ratio $K^+/\pi^+$ over the ratio $K^-/\pi^-$ of particle yields in heavy-ion collisions at $\sqrt{s_{NN}}\sim 8$ GeV (equivalent to $E_{\rm lab}\sim 30$ AGeV in fixed target experiments) called the "horn" effect \cite{Gazdzicki:1998vd})\footnote{See \cite{Cleymans:2004hj} for the references to the experimental data and an early attempt to explain the location of the "horn" within a statistical model.} may serve as such a phenomenon.
In the present paper $K^+/\pi^+$ and  $K^-/\pi^-$ ratios are investigated in a 2+1 flavor PNJL model on the base of phase shifts for pseudoscalar meson correlations (bound and scattering states) in the Beth-Uhlenbeck approach \cite{Hufner:1994ma,Zhuang:1994dw,Blaschke:2013zaa}.
\section{The $2+1$ flavor PNJL model}
\label{sec:partition}

We use a  $2+1$ flavor  NJL model with scalar and  pseudoscalar meson spectrum 
generalized by coupling to the Polyakov loop \cite{Hansen:2006ee,Costa:2008dp}
\begin{eqnarray}
\label{lagr}
{\cal L} &=& \bar{q} \left( i \gamma^\mu D_\mu + \hat{m}_0\right) q +
G_S \sum_{a=0}^{8} \left[ \left( \bar{q} \lambda^a q\right)^2+
\left( \bar{q} i \gamma_5 \lambda^a q\right)^2
\right]
-\mathcal{U}\left(\Phi[A],\bar\Phi[A];T\right). 
\end{eqnarray}
Here $q$ denotes the quark field with three flavors, 
$f=u,d,s$, and three colors, $N_c=3$; $\lambda^a$ are the flavor SU$_f$(3) 
Gell-Mann matrices ($a=0,1,2,\ldots,8$), $G_S$ is a coupling constant, $\hat{m}_0 = \mbox{diag}(m_{0,u}, m_{0,d}, m_{0,s})$ 
is the diagonal matrix of current quark masses which induces an explicit breaking of the otherwise global chiral symmetry of the Lagrangian (\ref{lagr}). 
The covariant derivative is defined as $D^{\mu}=\partial^\mu-i A^\mu$, with $A^\mu=\delta^{\mu}_{0}A^0$ 
(Polyakov gauge); in Euclidean notation $A^0 = -iA_4$.  
The strong coupling constant $g_s$ is absorbed in the definition of 
$A^\mu(x) = g_s {\cal A}^\mu_a(x)\frac{\lambda_a}{2}$, where 
${\cal A}^\mu_a$ is the (SU$_c$(3)) gauge field and $\lambda_a$ are the 
Gell-Mann matrices in SU$_c$(3) color space.

The effective potential for the (complex) $\Phi$ field was chosen in 
the polynomial form with the parametrization proposed in Ref.~\cite{Ratti:2005jh}:
\begin{eqnarray}\label{effpot}
&&\frac{\mathcal{U}\left(\Phi,\bar\Phi;T\right)}{T^4}=-\frac{b_2\left(T\right)}{2}\bar\Phi \Phi-
\frac{b_3}{6}\left(\Phi^3+ {\bar\Phi}^3\right)+
\frac{b_4}{4}\left(\bar\Phi \Phi\right)^2,\ \nonumber  \\ \label{Ueff}
&&b_2\left(T\right) = a_0+a_1\left(\frac{T_0}{T}\right)+a_2\left(\frac{T_0}{T}
\right)^2+a_3\left(\frac{T_0}{T}\right)^3~, \nonumber
\end{eqnarray}
where\footnote{Here we do not rescale the $T_0$ parameter of Polyakov loop potential.} $T_0 = 0.27$ GeV, $a_0 =6.75$, $a_1 = -1.95$, $a_2 = 2.625$, $a_3 =-7.44$, $b_3 =0.75$, $b_4 =7.5$.

To obtain the equation for the order parameters, one needs to minimize the grand thermodynamic potential 
$\Omega = -T \ln Z$, $Z=\int D\bar q D q \exp[\int dx {\cal L}]$ with respect to a variation of the parameters:
\begin{equation}
\frac{\partial \Omega}{\partial\langle\bar{q}q\rangle} = 0, \ \ \ \frac{\partial \Omega}{\partial \Phi} = 0, \ \ \ \frac{\partial \Omega}{\partial \bar{\Phi}} = 0.
\end{equation}
The quark masses $m_f$ are found by solving the gap equations
\begin{eqnarray}
m_f = m_{0,f} + 16\, m_f G_S I_1^f(T,\mu),
\end{eqnarray}
where integral $I_1^f(T,\mu)$ for finite temperature and chemical potential is defined as 
\begin{eqnarray}
\label{firstt}
I_1^f(T,\mu_f) &=& \frac{N_c}{4\pi^2} \int_0^\Lambda \frac{dp \, p^2}{E_f} \left(n^-_f - n^+_f \right).
\label{int_I1}
\end{eqnarray}
The generalized fermion distribution functions $n_f^{\pm}=f^+_\Phi(\pm E_f)$  \cite{Costa:2008dp, Blaschke:2014zsa} for quarks of flavor $f$ with positive (negative) energies in the presence of the Polyakov loop values $\Phi$ and $\bar{\Phi}$ are: 
\begin{eqnarray}
f^+_\Phi(E_f)=
\frac{(\bar{\Phi}+2{\Phi}Y)Y+Y^3}{1+3(\bar{\Phi}+{\Phi}Y)Y+Y^3}
~,
f^-_\Phi(E_f)=
\frac{({\Phi}+2\bar{\Phi}\bar{Y})\bar{Y}+\bar{Y}^3}{1+3({\Phi}+\bar{\Phi}\bar{Y})\bar{Y}+\bar{Y}^3}
~,
\label{f-Phi-bar}
\end{eqnarray}
where the abbreviations $Y={\rm e}^{-(E_f-\mu_f)/T}$ and $\bar{Y}={\rm e}^{-(E_f+\mu_f)/T}$ are used. The functions  (\ref{f-Phi-bar}) fulfil the relationship  $f^+_\Phi(-E_f)=1-f^-_\Phi(E_f)$, 
and they go over to the ordinary Fermi functions for $\Phi=\bar \Phi=1$.

The parameters used for the numerical studies in this work are  the bare quark masses  
$m_{0(u,d)}= 5.5~$MeV and  $m_{0s}= 138.6~$MeV, the three-momentum cut-off  
$\Lambda= 602~$MeV and the scalar coupling constant $G_{S}\Lambda^{2}= 2.317$. 
With these parameters one finds in the vacuum a constituent quark mass for light quarks of 367 MeV, a pion mass of 135 MeV and a pion decay constant $f_{\pi}=92.4~$MeV.

%

\begin{figure}[t]
	\centerline{
		\includegraphics[width=0.65\textwidth]{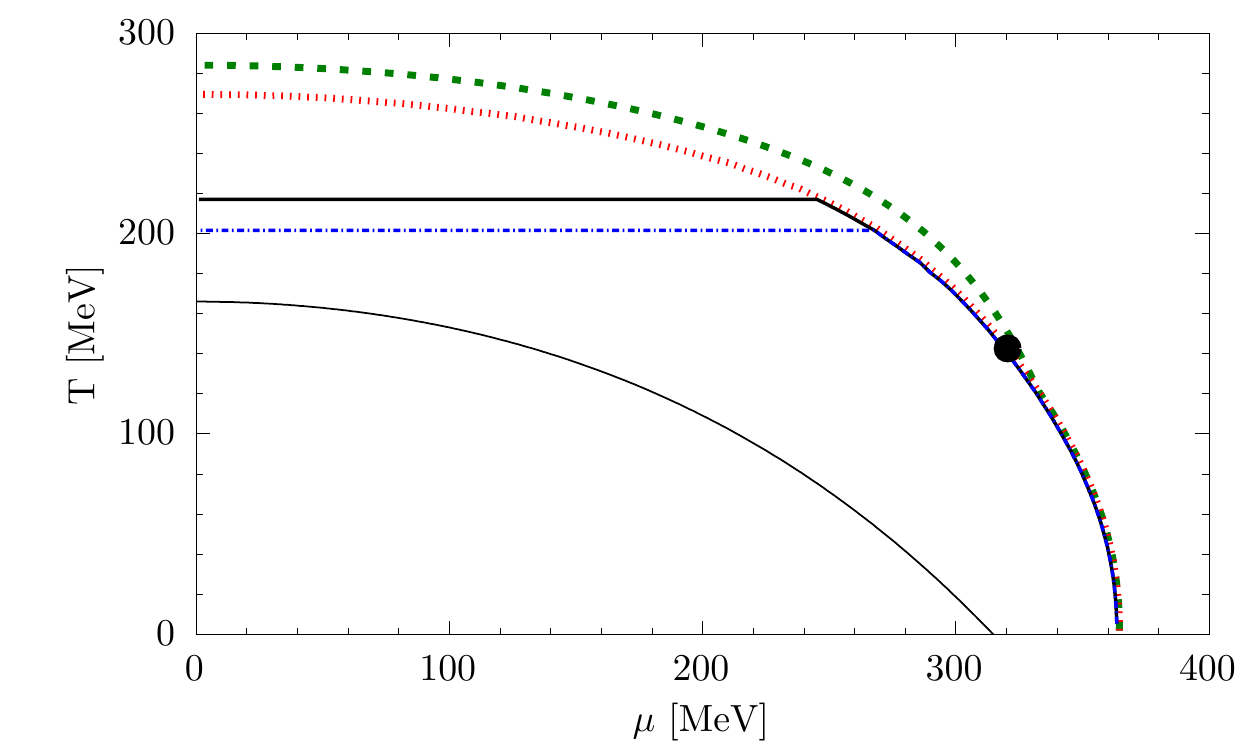}
	}
	\caption{
		Phase diagram of the PNJL model and lines of scan for $K/\pi$ ratio. The red dotted solid line corresponds to the first order phase transition or crossover transition and the black point denotes the CEP. The green dashed line is the Mott temperature for the pion, lines of scan of the $K^-\pi^-$ and $K^+/\pi^+$  ratios in T-$\mu_q$ plane are: chemical freezout line (thin black), critical line (red dotted), critical line+straight part for correct limit for $\mu_\pi=100$ MeV (blue dash-dotted)  and 
		for $\mu_\pi=134.5$ MeV (thick black). 
	}
	\label{Fig:LinesOfScan}
\end{figure}

In Fig.~\ref{Fig:LinesOfScan} we show the phase diagram of the present model.
To this end we find the positions of the minima of the temperature derivative (the steepest descent) of the quark mass as the chiral order parameter $dM/dT$ in the $T-\mu$ plane. 
These pseudocritical temperatures go over to the critical temperatures of the first order phase transition characterized by a jump of the quark mass at the corresponding position in the $T-\mu$ plane.

A characteristic feature of the phase diagram is that lowering the ratio $T/\mu \to 0$, the phase transition turns from crossover to first order. The chiral restoration is a result of the phase space occupation due to Pauli blocking which effectively reduces the interaction strength in the  gap equation.
The coupling to the Polyakov loop reduces the occupation of the phase space by quarks and therefore the pseudocritical temperatures are higher than in the corresponding NJL model.


\section{Mass spectrum for mesons at finite temperature and density}

The mass of the bound quark-antiquark state for the 2+1 flavor (P)NJL model can be defined from the pole condition of the meson propagator (the Bethe-Salpeter equation):
\begin{equation}
[\mathcal{S}^{M^a}_{f f'}(M_{M^a};\bar{0})]^{-1} =(2 G_s)^{-1} - \Pi_{f f'}^{M^a}(M_{M^a}+i\eta;\bar{0})=0,
\end{equation}
where the polarisation operator $\Pi^{M^a}_{ff'}$ is defined as
\begin{eqnarray}
\Pi^{M^a}_{ff'} (q_0 ,{\bf q}) &=& 2N_cT \sum_{n} \int \frac{d^3p}{(2\pi)^3}\mbox{tr}_{D}
\left[ S_{f}(p_n ,{\bf p}) \Gamma^{M^a}_{ff'} S_{f'}(p_n + q_0,{\bf p+q}) \Gamma^{M^a}_{ff'} \right],
\end{eqnarray} 
with $\Gamma^{S^a}_{ff'} = T^a_{ff'}$ for scalar meson and $\Gamma^{P^a}_{ff'}= i\gamma_5\ T^a_{ff'}$ for pseudo-scalar meson,
\begin{eqnarray}
T^a_{ff'}&=&
\left\{
\begin{array}{l}
(\lambda_3)_{ff'}, \\
(\lambda_1\pm i \lambda_2)_{ff'}/\sqrt{2}, \\
(\lambda_4\pm i \lambda_5)_{ff'}/\sqrt{2}, \\
(\lambda_6\pm i \lambda_7)_{ff'}/\sqrt{2}.
\end{array}
\right.
\end{eqnarray}

The matrix elements of the polarization operator can be represented in terms of two integrals which after summation over the Matsubara frequencies for mesons at rest in the medium (${\bf q}={\bf 0}$) are given by 
\begin{eqnarray}
&&\Pi^{P^a,S^a}_{ff'} (q_0+i\eta, {\bf 0}) = 4 \bigl\{  I_{1}^f(T,\mu_{f})+I_{1}^{f'}(T,\mu_{f'}) \mp \left[  (q_0+\mu_{ff'})^2 -(m_f \mp m_{f'})^2 \right] I_2^{ff'}(z,T,\mu_{ff'}) \bigr\}\, , \nonumber
\end{eqnarray}
where $\mu_{ff'}=\mu_{f}-\mu_{f'}$. 
The integral $I^f_1$ for finite temperature and chemical potential is given in Eq. (\ref{int_I1}) and $I^{ff'}_2$ has the 
following form
\begin{eqnarray}\label{int_I2}
I_2^{ff'} (z,T,\mu_{ff'}) &=& \frac{N_c}{4\pi^2} \int_0^\Lambda \frac{dp \, p^2}{E_fE_{f'}}
\Biggl[ \frac{E_{f'}}{(z-E_f-\mu_{ff'})^2-E_{f'}^2} \,\, n^-_f 
-  \frac{E_{f'}}{(z+E_f-\mu_{ff'})^2-E_{f'}^2} \,\, n^+_f
\nonumber \\ 
&& +\frac{E_f}{(z+E_{f'}-\mu_{ff'})^2-E_f^2} \,\,  n^-_{f'} 
- \frac{E_f}{(z-E_{f'}-\mu_{ff'})^2-E_f^2} \,\,  n^+_{f'} 
\Biggr], 
\end{eqnarray}
with $E_{f}=\sqrt{{p}^2+m_{f}^2}$ being the quark energy dispersion relation. 

This method works well as long as the particle is a true bound state, that is below  the Mott temperature ($T_{\rm Mott}^{M}$), while $q_0=M_{M^a}< m_{{\rm thr}, ff'}$,  with $m_{{\rm thr}, ff'} = m_f+m_{f'}$. 
Then above the Mott temperature for $M_{M^a}>m_{{\rm thr},ff'}$ the meson becomes an unbound state. 

To describe the mesonic states in dense matter it is preferable to use the phase shift of the quark-antiquark correlation in the considered mesonic interaction channel. 
The meson propagator can be rewritten in the ''polar'' representation:
\begin{equation}
\mathcal{S}^{M^a}_{f f'}(\omega,\bar{q})= |\mathcal{S}^{M^a}_{f f'}(\omega,\bar{q})|e^{\delta_M(\omega,\bar{q})}
\end{equation}
with meson phase shift 
\begin{align}
\delta_M(\omega,\bar{q}) = -{\rm{arctan}}\left\{\frac{{\rm{Im}}([{\mathcal{S}}^{M^a}_{f f'}(\omega-i\eta,\bar{q})]^{-1})}{{\rm{Re}}([\mathcal{S}^{M^a}_{f f'}(\omega+i\eta,\bar{q})]^{-1})}\right\}.
\label{phaseshift}
\end{align}
To define the mass we determine the energy $\omega$ where the phase shift assumes the value $\pi$. In the rest frame of the meson this energy corresponds to the mass.  Below the Mott temperature, the phase shift jumps from zero to $\pi$ at this position so that its derivative  is a delta function, characteristic for a true bound state. Then the Beth-Uhlenbeck formula for the mesonic pressure can be given the following form:
\begin{eqnarray}
P_{\rm M} = d_{\rm M}\int\frac{\rm{d}^3 q}{(2\pi)^3} \int_0^\infty\frac{{\rm d}\omega}{(2\pi)}\delta_M(\omega,\bar{q})g(\omega\pm\mu_M),
\label{BUU}
\end{eqnarray}
where $g(E) = (e^{E/T}-1)^{-1}$ is the Bose function. 
One can simplify expression under the assumption that, even in the medium, the phase shifts are
Lorentz invariant and depend on $\omega$ and $q$ only via the Mandelstam variable $s=\omega^2-q^2$ in the form 
$\delta_M(\omega, \bar{q})=\delta_M(\sqrt{s}, \bar{q} = 0)\equiv \delta_M(\sqrt{s}; T, \mu_M)$ for given temperature and chemical potential of the medium. Then the pressure can be rewritten as:
\begin{eqnarray}
P_{\rm M} &=& d_{\rm M} \int_0^\infty\frac{{\rm d}\mathcal{M}}{2\pi}\delta_M(\mathcal{M})\int\frac{\rm{d}^3 q}{(2\pi)^3}\frac{\mathcal{M}}{E_M}g(E_M\pm\mu_{\rm M}), \label{BUU}
\end{eqnarray}
where $E_M = \sqrt{\mathcal{M}^2+q^2}$. 
In the case when the continuum of the scattering states can be neglected, that is when it is separated by a sufficiently large energy gap from the bound state, we obtain as a limiting case the thermodynamics of a statistical ensemble of on-shell correlations (resonance gas).

\begin{figure}[t]
	\centerline{
		\includegraphics[width=0.33\textwidth]{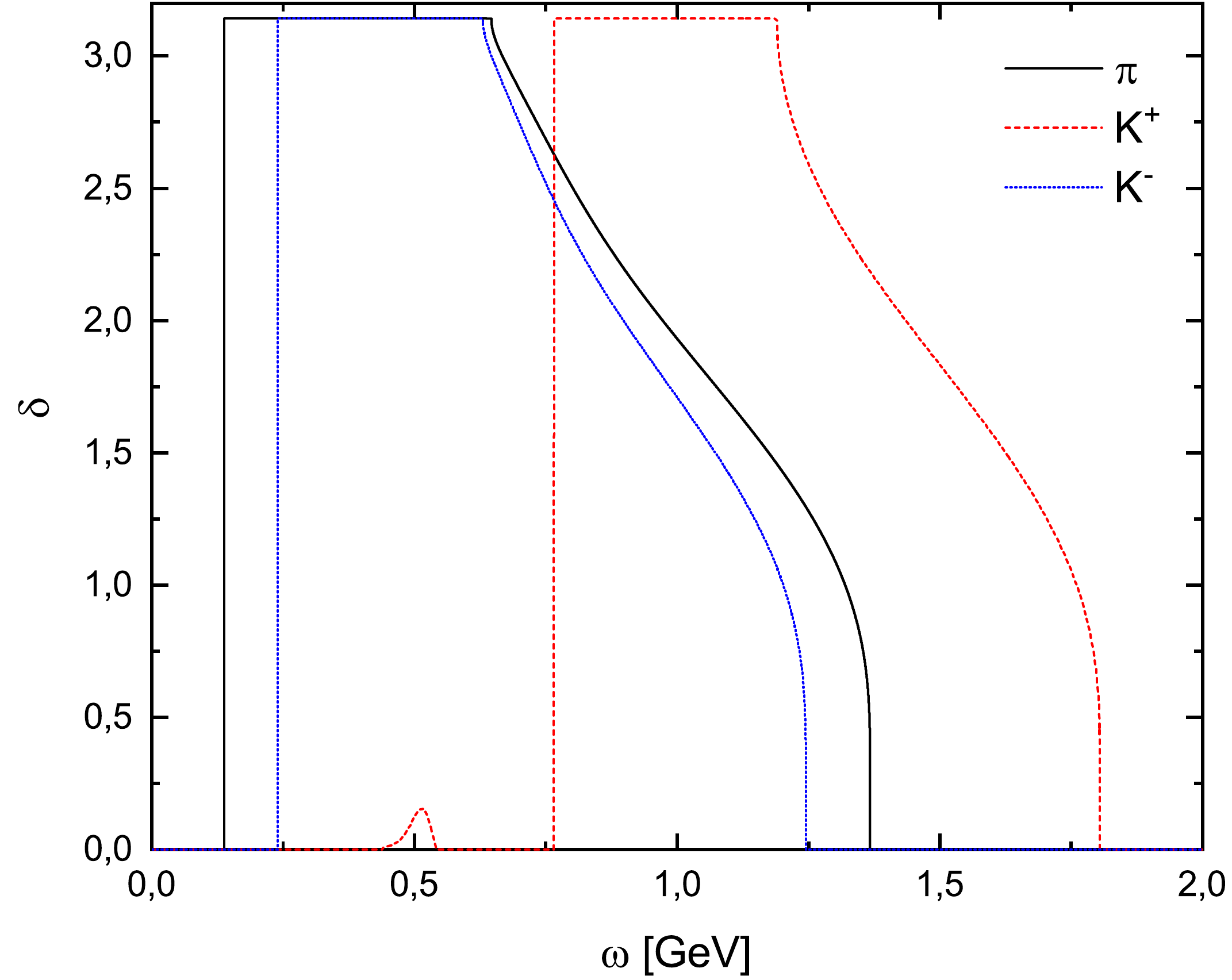}
		\includegraphics[width=0.33\textwidth]{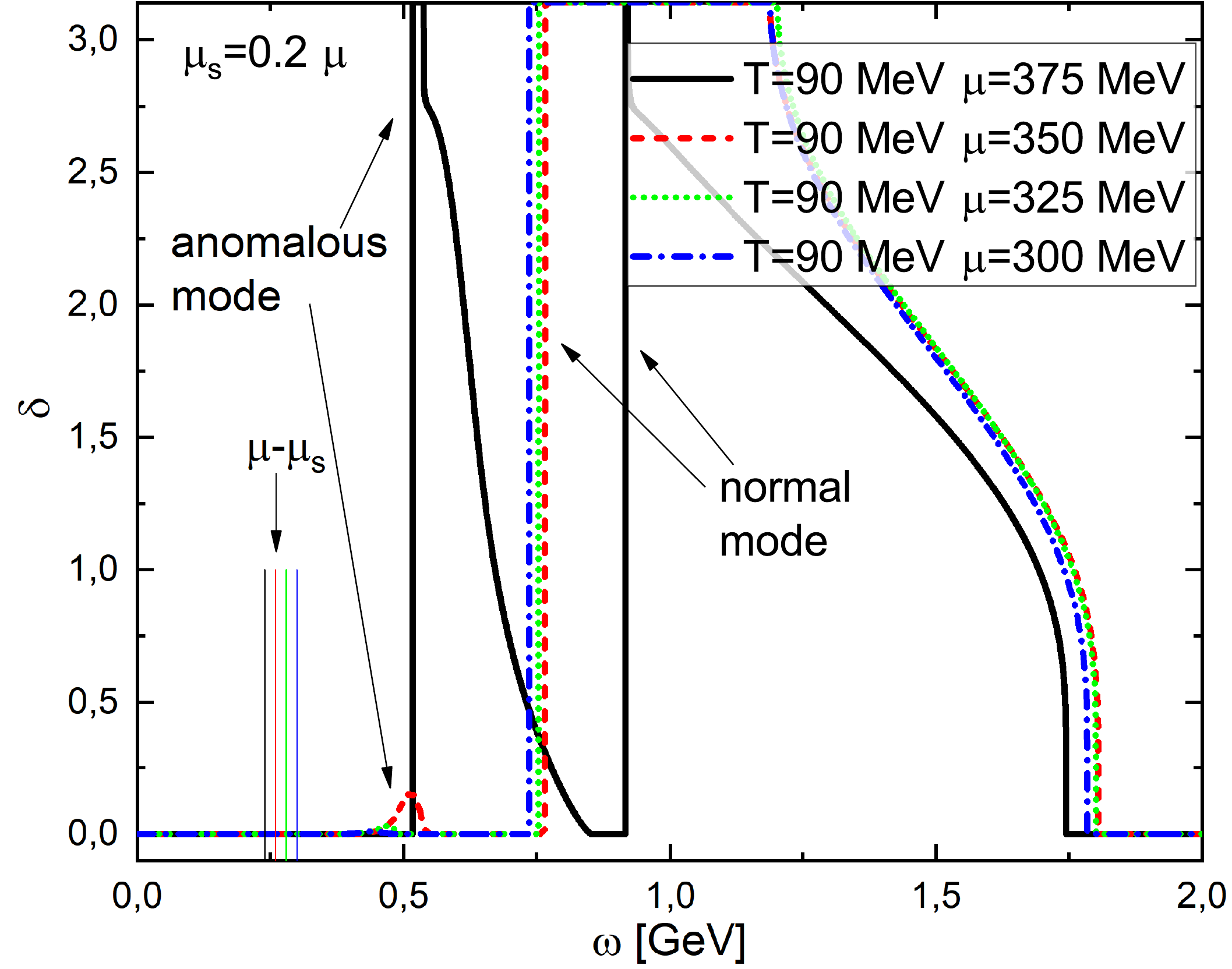}
		\includegraphics[width=0.33\textwidth]{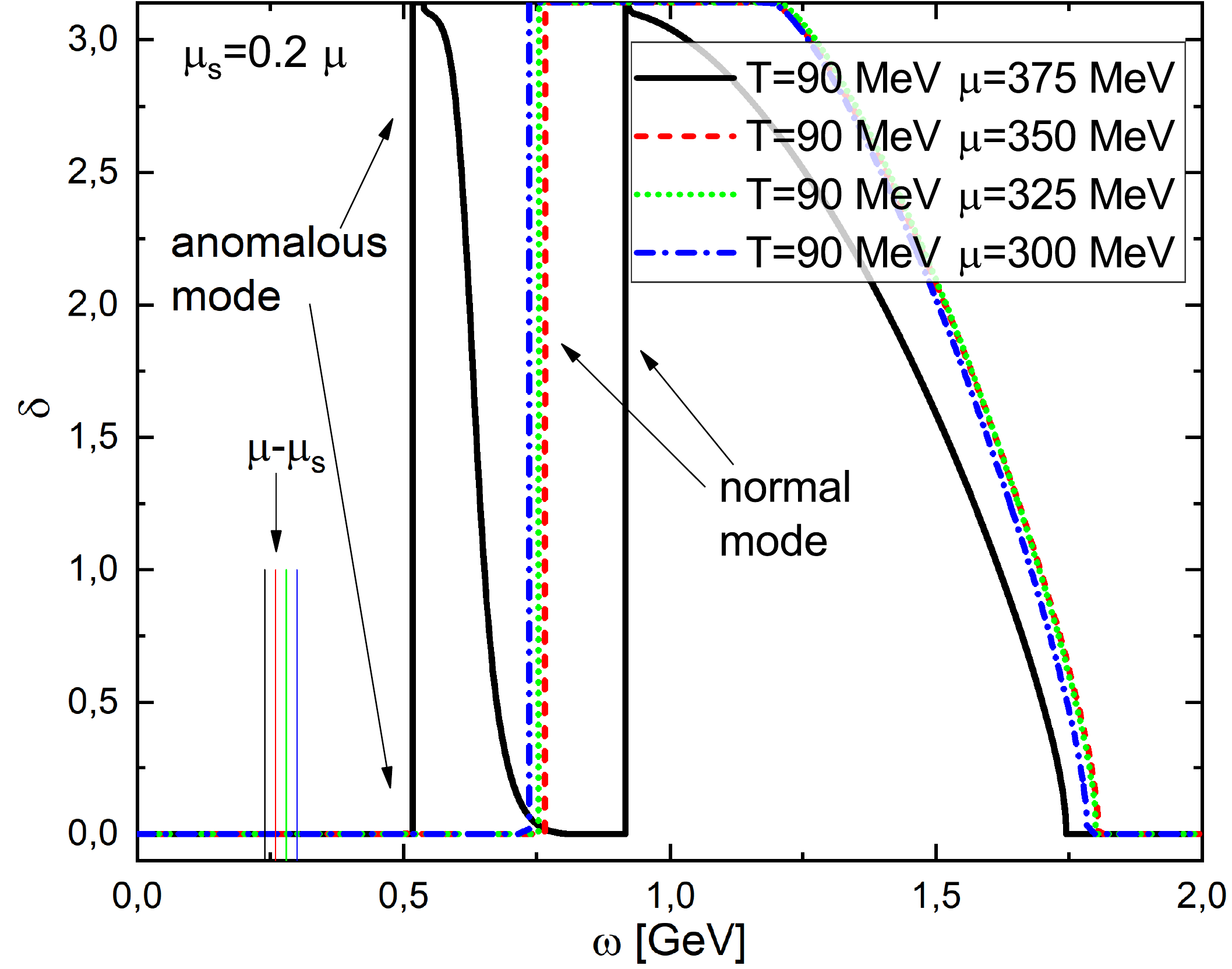}
}
	\caption{Dependence of the phase shift in the pion, $K^+$ and $K^-$ meson channels on the center of mass energy for $T= 90$ MeV and nonstrange chemical potential $\mu=350$ MeV (left panel), dependence of the phase shift in the $K^+$  channel for $T= 90$ MeV  and nonstrange chemical potential $\mu=300,325,350,375$ MeV (central panel), dependence of the combination of phase shift $ \delta_{i}(M) - \sin(2\delta_{i}(M))/2$ of Eq.~(\ref{GenBUUReplacemet}) in the generalized Beth-Uhlenbeck approach in the $K^+$  channel for $T= 90$ MeV  and nonstrange chemical potential $\mu=300,325,350,375$ MeV (right panel) .
	}
	\label{Fig:ShiftsPiKaMiKaPl}%
\end{figure}

In Fig.~\ref{Fig:ShiftsPiKaMiKaPl} we show the behaviour of phase shifts for the pion and for charged kaons. 
In left panel of the figure these phase shifts are shown  for $T= 90$ MeV  and chemical potential $\mu=300$ MeV and  strange chemical potential is always set in our calculations to $\mu_s=0.2\mu$.
For these $T$ and $\mu$ the phase shift of the pion is almost the vacuum one while for charged kaons there is a splitting of  masses and additionally a small anomalous mode appears for $K^+$. 
In the central and right panels of the figure the phase shift $ \delta(M)$ in the $K^+$  channel  and its combination 
$ \delta(M) - \sin(2\delta(M))/2$ for the generalized Beth-Uhlenbeck approach are shown for $T= 90$ MeV  and chemical potentials $\mu=300,325,350,375$ MeV.

For $\mu=0$ their modification with increasing $T$ is similar to that of the pions, with the gap between the bound state and the continuum diminishing with temperature and becoming zero for $T=300$ MeV, above the kaon Mott temperature, where the kaon becomes a resonance in the continuum. 

As can be seen from Fig.~\ref{Fig:ShiftsPiKaMiKaPl}, a finite chemical potential 
removes the degeneracy of the meson masses in the strange channels and a mass difference arises between $K^{+}$ and $K^{-}$. The chemical potential shifts the pole in the propagator, which results in a reduction of the pseudocritical temperature $T_{c}$ and therefore also in a reduction of the meson Mott temperatures $T_{\rm Mott}^{M}$. 

At nonzero chemical potential and low $T$, the splitting of mass in charged multiplets is due to the excitation of the Dirac sea modified by the presence of the medium. 
In dense baryon matter the concentration of light quarks is very high \cite{Stachel_ss}. 
Therefore, the creation of a $s\bar{s}$ pair dominates because of the Pauli principle: when the Fermi energy for light quarks is higher than the $s\bar{s}$-mass, the creation of the latter is energy-efficient.  
The increase in the $K^+$  ($\bar{u}s$) mass, with respect to that of $K^-$ ($\bar{s}u$), is justified again by the Pauli blocking for s-quarks (see for discussion \cite{Lutz,Ruivo_1996,CostaKalin_2003,CostaKalin_2004}).
Technically, to describe the mesons in dense matter, it is needed to relate the chemical potential of quarks with Fermi momentum $\lambda_i$,  $\mu_i = \sqrt{\lambda_i^2 + m_i^2}$. 
The latter affects  the limits of integration in Eqs.~(\ref{int_I1}) and (\ref{int_I2}). 
It is obvious that the pion for the chosen cases ($m_u = m_d$) is still degenerate.


\section{Kaon to pion ratio}
From Eq.~(\ref{BUU}) the meson partial number density as off-shell generalization of the number density of the bosonic species 
$M$ is
\begin{eqnarray}
n_M(T)&=& 
d_M\int\frac{\rm{ d}^3 q}{(2\pi)^3} \int \frac{d\mathcal{M}}{2\pi} g_M(E_M-\mu_M)\frac{d\delta_{M}(\mathcal{M})}{d\mathcal{M}}
\nonumber \\
&=& \frac{d_{\rm M}}{T} \int_0^\infty\frac{{\rm d}\mathcal{M}}{(2\pi)}\delta_M(\mathcal{M})\int\frac{\rm{ d}^3 q}{(2\pi)^3}g(E_M - \mu_{\rm M})(1+g(E_M - \mu_{\rm M})).
\label{BUUpartial}
\end{eqnarray}

%
%
The chemical potential for kaons can be defined (see for example \cite{naskret,pot_K}) from 
$\mu_{\rm M} = B_q\mu_B + S_q\mu_s + I_q\mu_q$, and in the isospin   symmetry case ($I_q = 0$), the result is 
$\mu_K =\mu_u-\mu_s$. 
The chemical potential for pions also is a phenomenological parameter, but it has its origin in the nonequilibrium nature of the distribution function of the pions for which, in contrast to the equilibrium case, the pion number is a quasi conserved quantity,
see also \cite{Gavin:1990up}.
In the works  \cite{pot_pi,BegunPRC90,naskret},  for example,  it was chosen as a constant, $\mu_\pi = 135$ MeV. 
In \cite{BegunPRC90} it was supposed that $\mu_\pi$ can depend on T.  

\begin{figure}[!th]
	\centerline{
		\includegraphics[ width=0.45\textwidth]{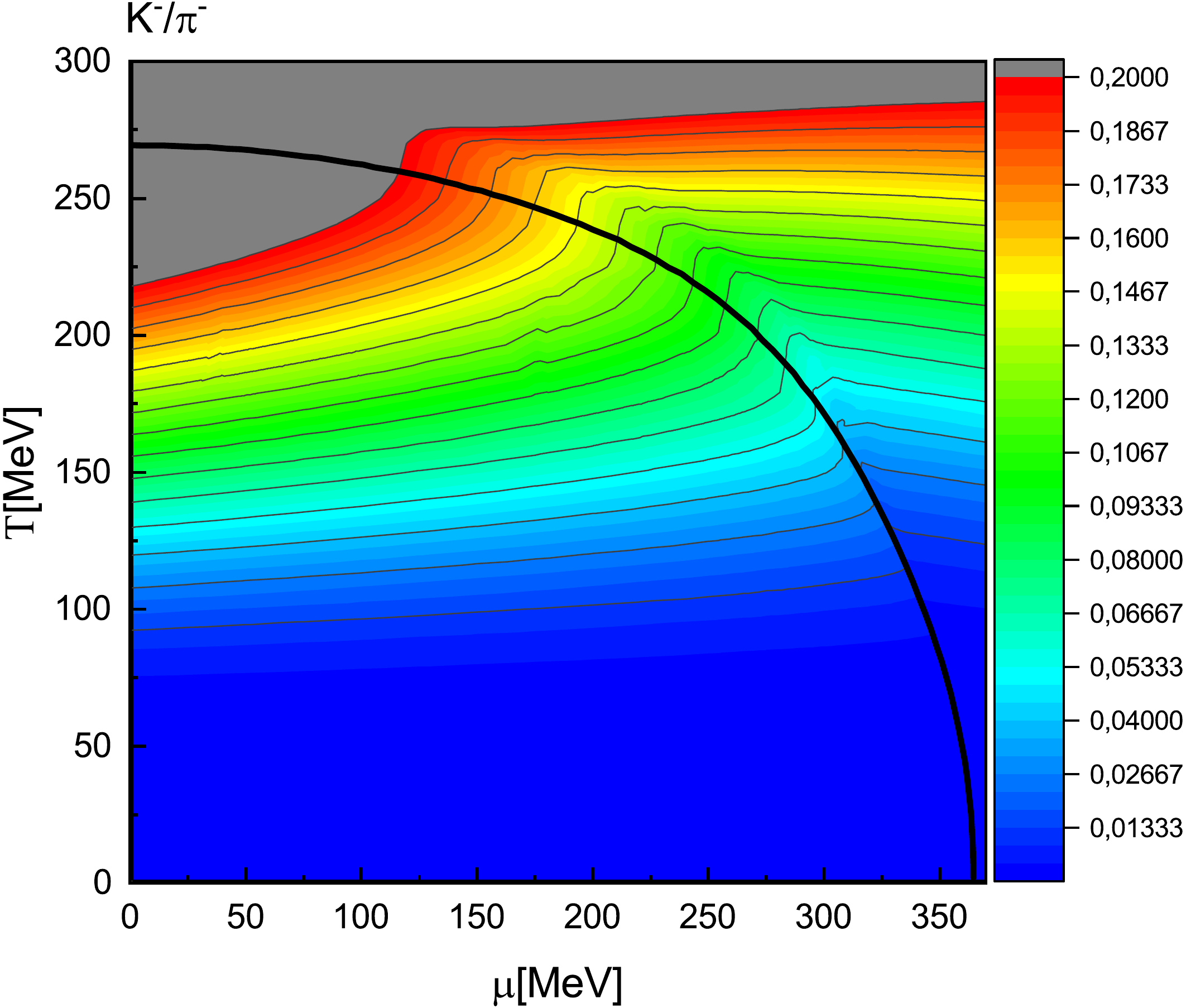}
		\includegraphics[width=0.45\textwidth]{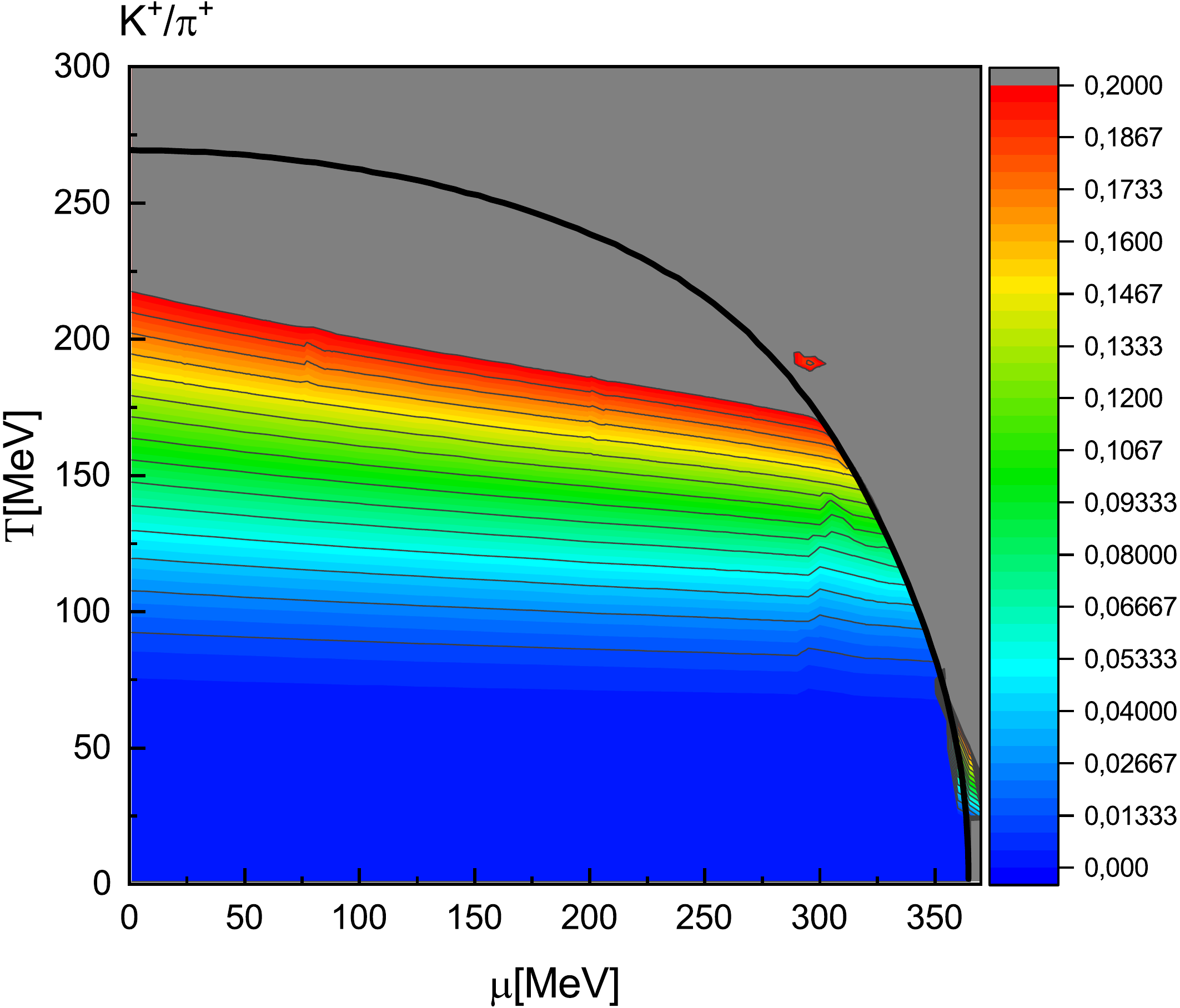}
	}
	\centerline{
		\includegraphics[ width=0.45\textwidth]{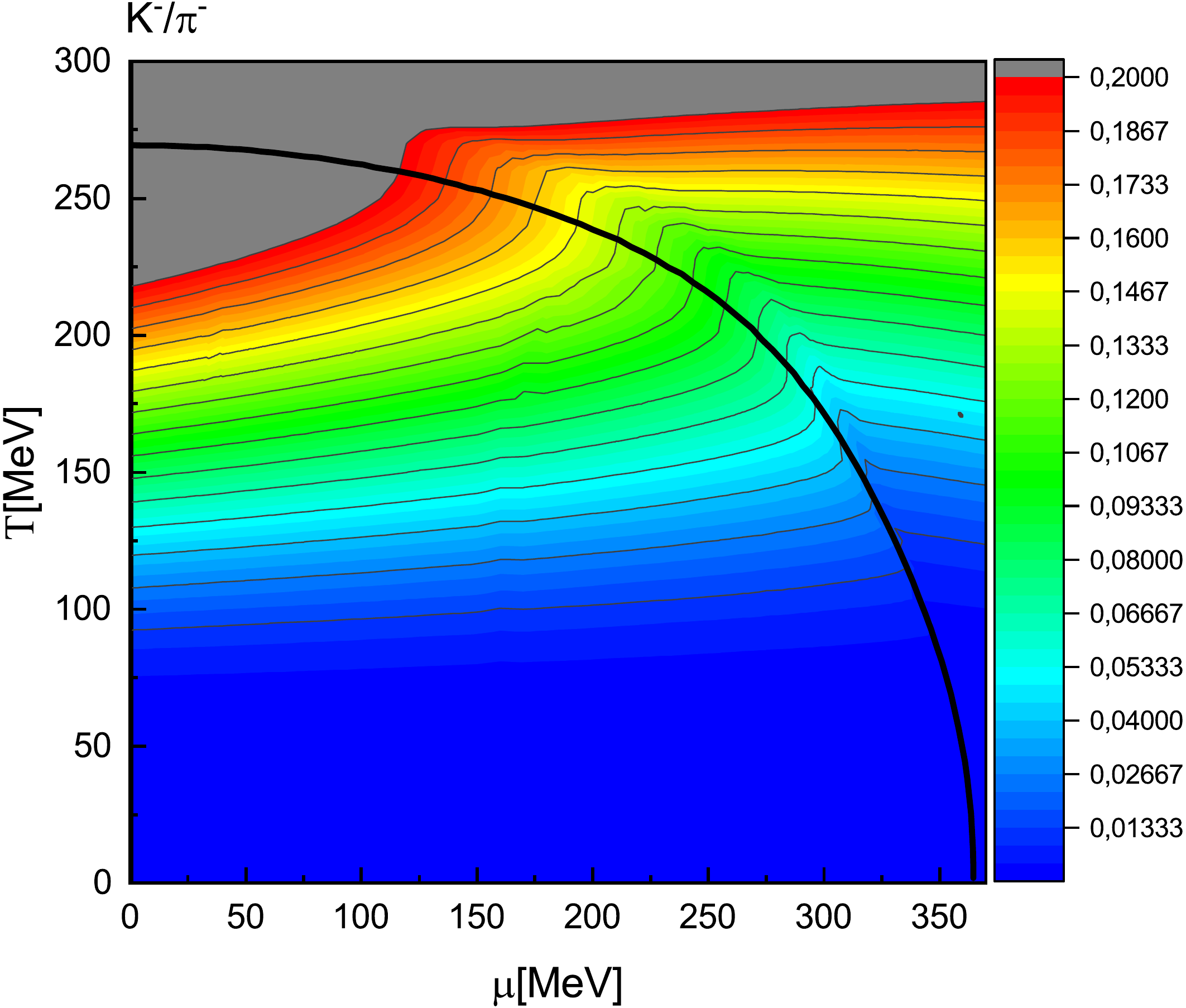}
		\includegraphics[width=0.45\textwidth]{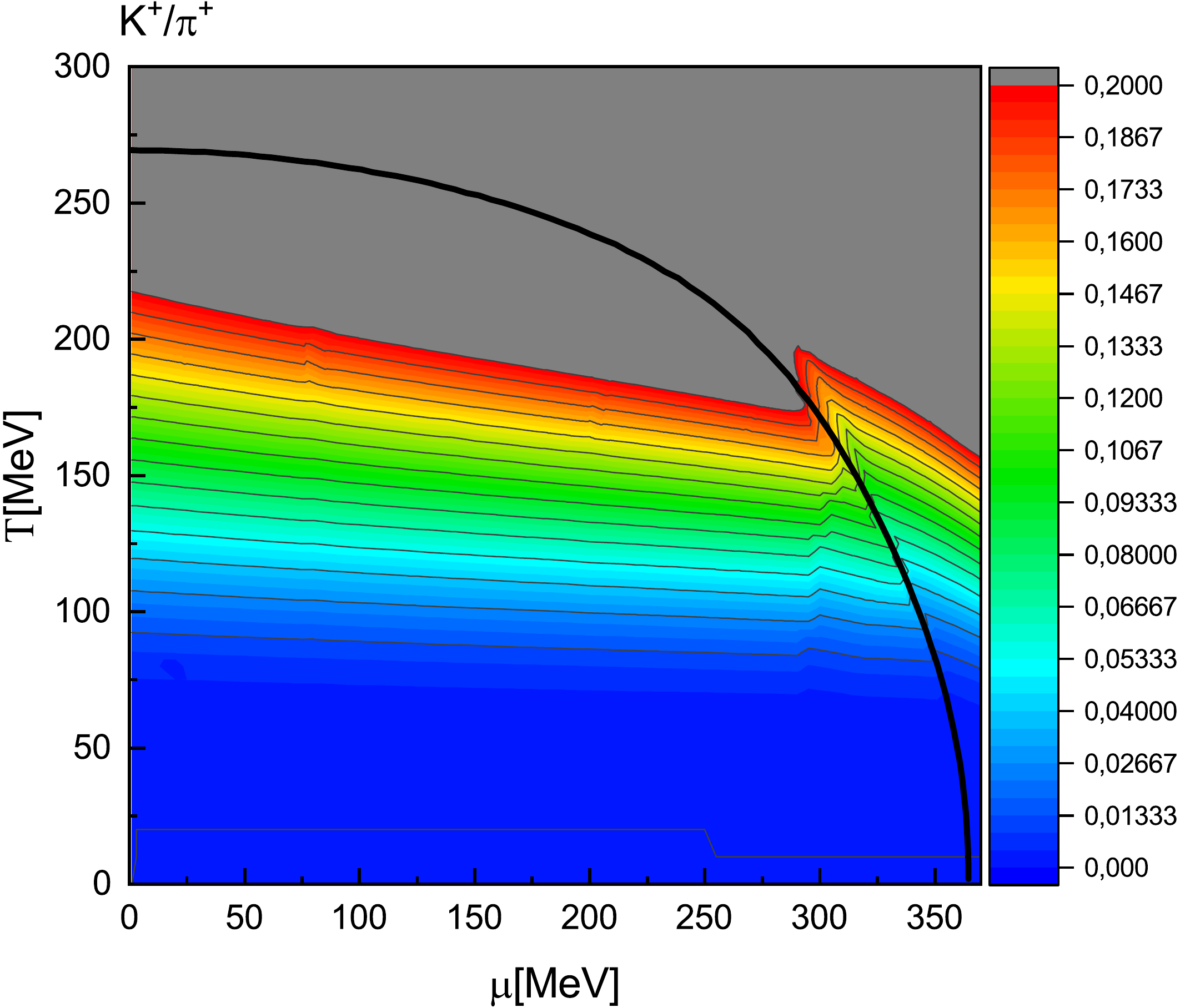}
	}
	\caption{
		The $K^-/\pi^-$ and $K^+/\pi^+$  ratio in the T-$\mu_q$ plane. The two upper panels correspond to full phase shifts while the two lower panels show phase shifts when the anomalous mode is removed. 
		Only the interval of ratios relevant for a comparison with the experimental data is shown. 
	}
	\label{Fig:KaPlKaMiFullNoAnomalousMode}
\end{figure}

The Beth-Uhlenbeck expression for the ratio of the yields of kaons and pions is defined as ratio of their partial number densities 
\begin{eqnarray}
	\label{K+pi+n-ratio}
	\frac{n_{K^\pm}}{n_{\pi^\pm}} = 
	\frac{\int dM \int d^3p\  (M/E)g_{K^\pm}(E)[1+g_{K^\pm}(E)]\delta_{K^\pm}(M)}
	{\int dM \int d^3p\ (M/E)g_{\pi^\pm}(E)[1+g_{\pi^\pm}(E)]\delta_{\pi^\pm}(M)}\ , 
\end{eqnarray}
When comparing to the ratio of the partial pressures, we observe that the only difference is the Bose enhancement factor which shall be important at best for the pions.
	Note that for the generalized Beth-Uhlenbeck approach one should make in Eq.~(\ref{K+pi+n-ratio}) the replacement
\begin{eqnarray}
\label{genBUU}
\delta_{i}(M) \to \delta_{i}(M) - \sin(2\delta_{i}(M))/2.
\label{GenBUUReplacemet}
\end{eqnarray}

There are four cases for the definition of kaon to pion ratios: 
with the partial pressures \eqref{BUU} or the partial densities \eqref{BUUpartial} and with or without the generalized 
Beth-Uhlenbeck replacement \eqref{genBUU}. 
An additional question is the possible role of the anomalous mode for the kaon ratios.

We found that in principle all these cases produce a strong enhancement of the $K^+/\pi^+$ ratio over $K^-/\pi^-$ but they are sensitive to particular details like the pion chemical potential.  
For the comparison with experimental data we consider the case with partial densities \eqref{BUUpartial} with the generalized BU replacement \eqref{genBUU} and investigate the influence of the low energy anomalous mode for two values of the pion chemical potential, $\mu_\pi = 100$  and $134.5$ MeV.

The kaon to pion ratios are shown in Fig.~\ref{Fig:KaPlKaMiFullNoAnomalousMode} for the case of partial densities with generalized BU replacement for a pion chemical potential $\mu_\pi = 100$ MeV. 
The full result is shown together with those when in the phase shift the anomalous part is omitted.

\begin{figure}[!th]
	\centerline{
		\includegraphics[width=0.49\textwidth]{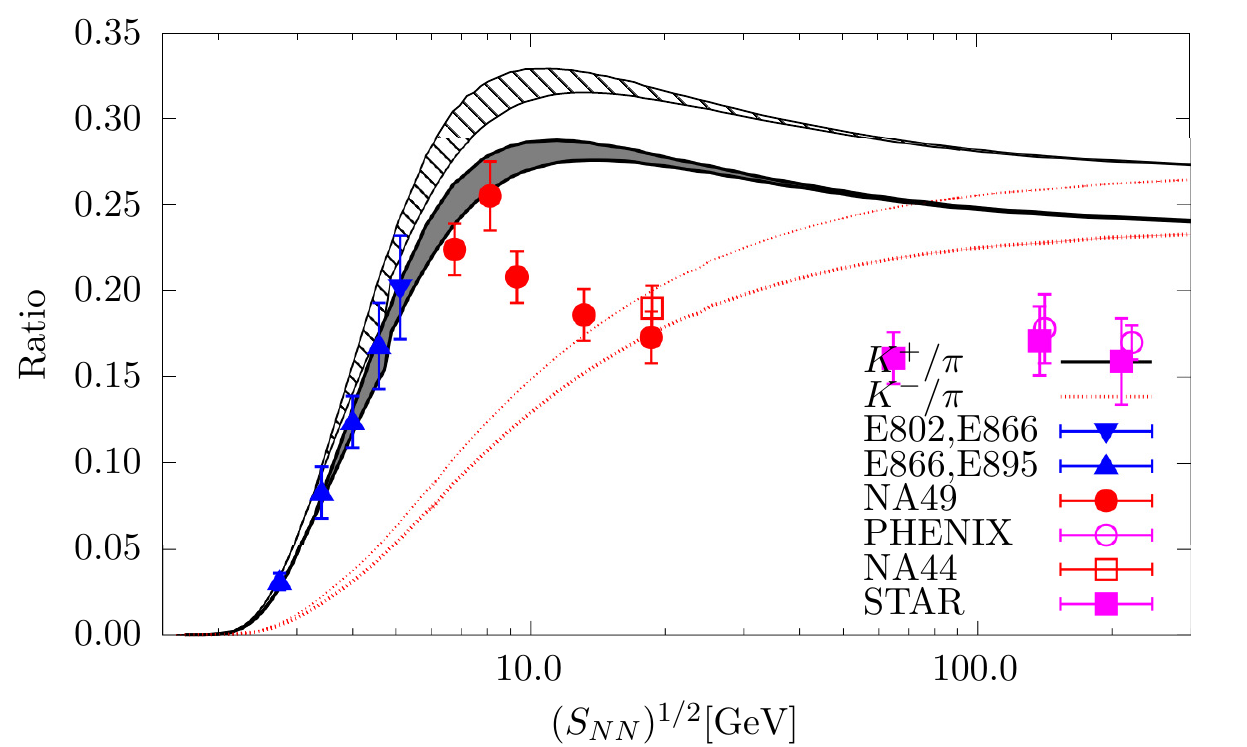}
		\includegraphics[width=0.49\textwidth]{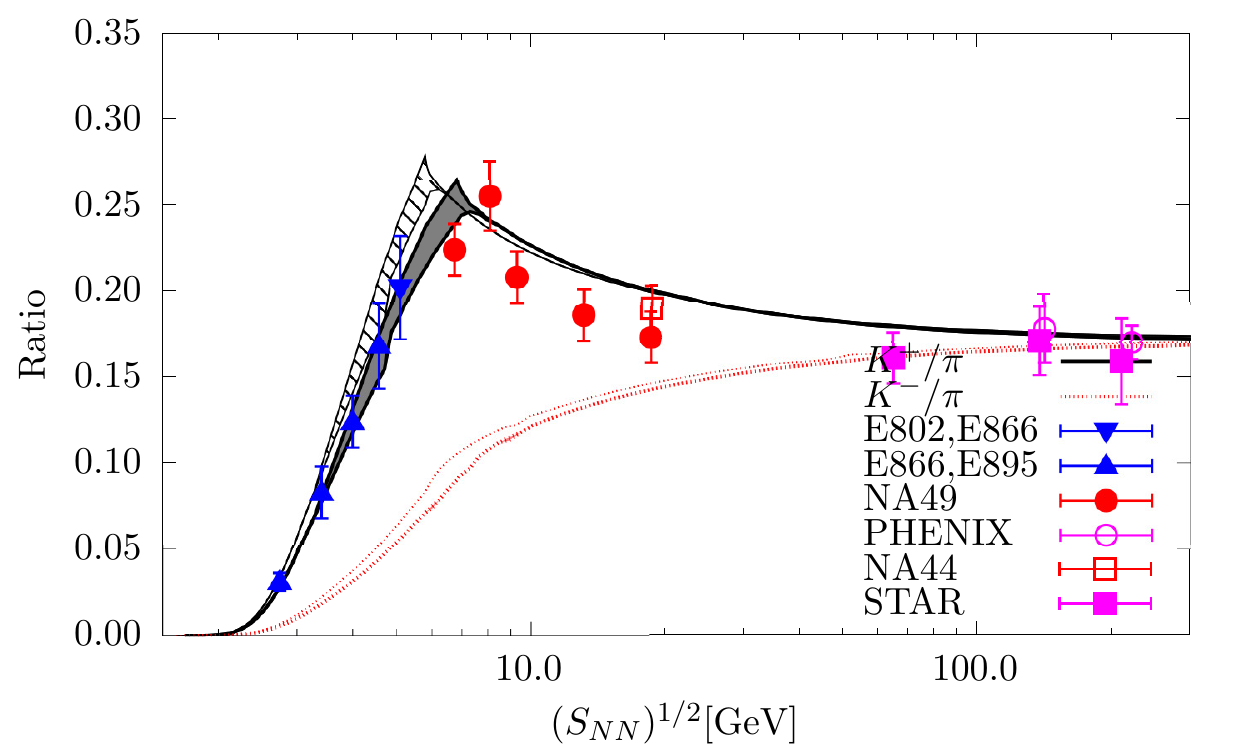}
	}
	\caption{
		The $K^-/\pi^-$ and $K^+/\pi^+$  ratios  compared to the experimental data for different scan lines  from Fig. \ref{Fig:LinesOfScan}.
	The left panel is a scan for the line near the phase transition, and the right panel is a scan by the phase transition and a straight constant temperature line.  
	The black line is the $K^+/\pi^+$ ratio and the red dotted one is $K^-/\pi^-$. 
	Thin lines correspond to the case of a pion chemical potential $\mu_\pi=100$ MeV and thick ones to $\mu_\pi=134.5$ MeV. 
	The shaded region between the lines for $K^+$ corresponds to the contribution of the anomalous mode. 
	}
	\label{Fig:Ratios}
\end{figure}

\section{Comparison with experimental data}

One can see from Fig. ~\ref{Fig:KaPlKaMiFullNoAnomalousMode} that for the $K^-/\pi^-$ ratio the influence of the anomalous mode is negligible. 
For $K^+/\pi^+$ the difference starts near the phase transition line and after the phase transition the anomalous mode has a strong influence on the ratio.

First we check the actual chemical freeze-out  line which leads to rather poor description of ratios in comparison with experiment.
The data is too low and there is no any trace of a horn for the $K^+/\pi^+$ ratio. 
This is due to fact that the chemical freeze-out curve lies far from phase transition line  in the PNJL model and therefore nothing could happen there apart from heating a gas of mesons. 

In order to relate the model results with the actual phenomenology of chemical freeze-out in heavy-ion collisions one can take a different scan region in order to check the influence of the model phase transition on the ratios. 
The problem is how to relate the model points with the experimental ones. 
Here we use the idea to map points with a fixed value of $\mu/T$ on the line in phase diagram of our PNJL model to points on the curve fitted to statistical model analyses \cite{Dubinin:2016wvt}.
Let us take this scan line as a critical line in PNJL model, see the left panel of Fig.~\ref{Fig:Ratios}. 
One can see an acceptable description for low energies but still for high $\sqrt{s_{NN}}$, i.e. the right side of the graph, the model overshoots the experimental data. 
This is due to fact that the pseudocritical temperature at zero chemical potential in the PNJL model is too high when compared with lattice QCD and with the fit of the freeze-out line. 
A simple solution is to somehow change the scan line in order to reproduce the limiting value for the $K^+/\pi^+$ ratio. The suggestion for this  line is simple: with decreasing $\mu/T$ after the "horn" the $K^+/\pi^+$ ratio should decrease and  $K^-/\pi^-$ should increase.  
From Fig.~\ref{Fig:KaPlKaMiFullNoAnomalousMode}  one can see that this line could be just a straight line at some fixed $T$ and could be fixed at zero $\mu$. 
These scan lines are shown in Fig.~\ref{Fig:LinesOfScan} and the results for the ratios are shown in the right panel of 
Fig.~\ref{Fig:Ratios}.
The contribution of anomalous mode is shown as shaded region in Fig.~\ref{Fig:Ratios} (the anomalous mode contribution is the main difference of the present paper with Ref.~\cite{Friesen:2018ojv}). 
One can see that the anomalous mode contributes visibly for the $K^+/\pi^+$ ratio for a scan near the phase transition region. For the points which are far from the phase transition the anomalous mode contribution is negligible for the $K^+/\pi^+$ ratio. 
For the $K^-/\pi^-$ ratio, the anomalous mode contribution is always negligible.

A new result of this work is the conjecture that the peak of the $K^+/\pi^+$ ratio may be related to the onset of Bose condensation for the pions as a consequence of the overpopulation of the pion phase space beyond a certain collision energy. 
This conjecture may be supported by the recent finding that the occurrence of the "horn" effect is strongly dependent on the  system size: while it is well pronounced in Pb+Pb collisions it is absent for Ar+Sc collisions \cite{Podlaski}. 

\funding{This work was supported by the Russian Fund for Basic Research under grant No. 18-02-40137.}
\conflictsofinterest{The authors declare no conflict of interest.} 
\reftitle{References}

\end{document}